\documentclass[a4paper,11pt]{article}
\pdfoutput=1 

\usepackage{jcappub} 

\usepackage[T1]{fontenc} 
\usepackage{multirow} 

\newcommand{\bq}{\boldsymbol q}
\newcommand{\bx}{\boldsymbol x}
\newcommand{\bk}{\textbf{k}}

\newcommand{\ihmpc}{\,h{\rm Mpc}^{-1}}

\newcommand{\change}[1]{\textcolor{black}{{{#1}}}}

\title{A comparison of template vs.\ direct model fitting for redshift-space distortions in BOSS}
\author[a]{Mark Maus}
\author[c]{Shi-Fan Chen}
\author[a,b]{Martin White} 
\emailAdd{mark.maus@berkeley.edu}

\affiliation[a]{Department of Physics, University of California, Berkeley, CA 94720, USA}
\affiliation[b]{Physics Division, Lawrence Berkeley National Laboratory, Berkeley, CA, USA}
\affiliation[c]{Institute for Advanced Study, 1 Einstein Drive, Princeton, NJ 08540, USA}
\abstract{The growth of large-scale structure, as revealed in the anisotropic of clustering of galaxies in the low redshift Universe \change{at z<2}, provides a stringent test of our cosmological model.  The strongest current constraints come from the BOSS and eBOSS surveys, with uncertainties on \change{$\sigma_8$, the amplitude of clustering on an 8 $h^{-1}$Mpc scale,} of less than 10 per cent.  A number of different approaches have been taken to fitting this signal, leading to \change{discrepancies of up to $1\sigma$ in the measurements of} the amplitude of fluctuations at late times.  We compare in some detail two of the leading approaches, one based on fitting a template cosmology whose amplitude and length scales are allowed to float with one based on \change{varying the underlying parameters of a cosmological model directly}, when fitting to the BOSS DR12 data.  Holding the input data, scale cuts, window functions and modeling framework fixed we are able to isolate the cause of the differences and discuss the implications for future surveys.  }

\begin{document}
\maketitle
\flushbottom

\section{Introduction}
\label{sec:intro}

The Universe we observe contains \change{structures} on essentially all scales that we can probe, from the largest superclusters to stellar systems.  All of \change{these structures are} believed to have arisen from quantum fluctuations in the primordial Universe that were amplified by gravitational instability \cite{Peacock99,Dodelson03,Baumann22}.  It thus contains information about astrophysics, cosmology and fundamental physics \cite{Amendola18}.  Of particular interest to us is the large-scale structure (LSS; on scales larger than a few Mpc), for which gravity is the dominant force.  Such LSS can be mapped in large galaxy redshift surveys within which the 3D \change{positions} of objects are determined by a combination of angular position and redshift, with the latter affected by both the Hubble flow and peculiar velocities. \change{The line-of-sight(LOS) component of these peculiar velocities affects the inferred distances of galaxies, introducing anisotropies in the clustering signal in LSS observations} \citep{Kaiser87,Hamilton92}. These so-called redshift-space distortions (RSD) present both a modeling challenge and additional information.  An accurate measurement of the growth of LSS through RSD is one of the key science goals of current and future galaxy redshift surveys \cite{DESI,Euclid}.

\begin{figure}
    \centering
    \resizebox{\textwidth}{!}{\includegraphics{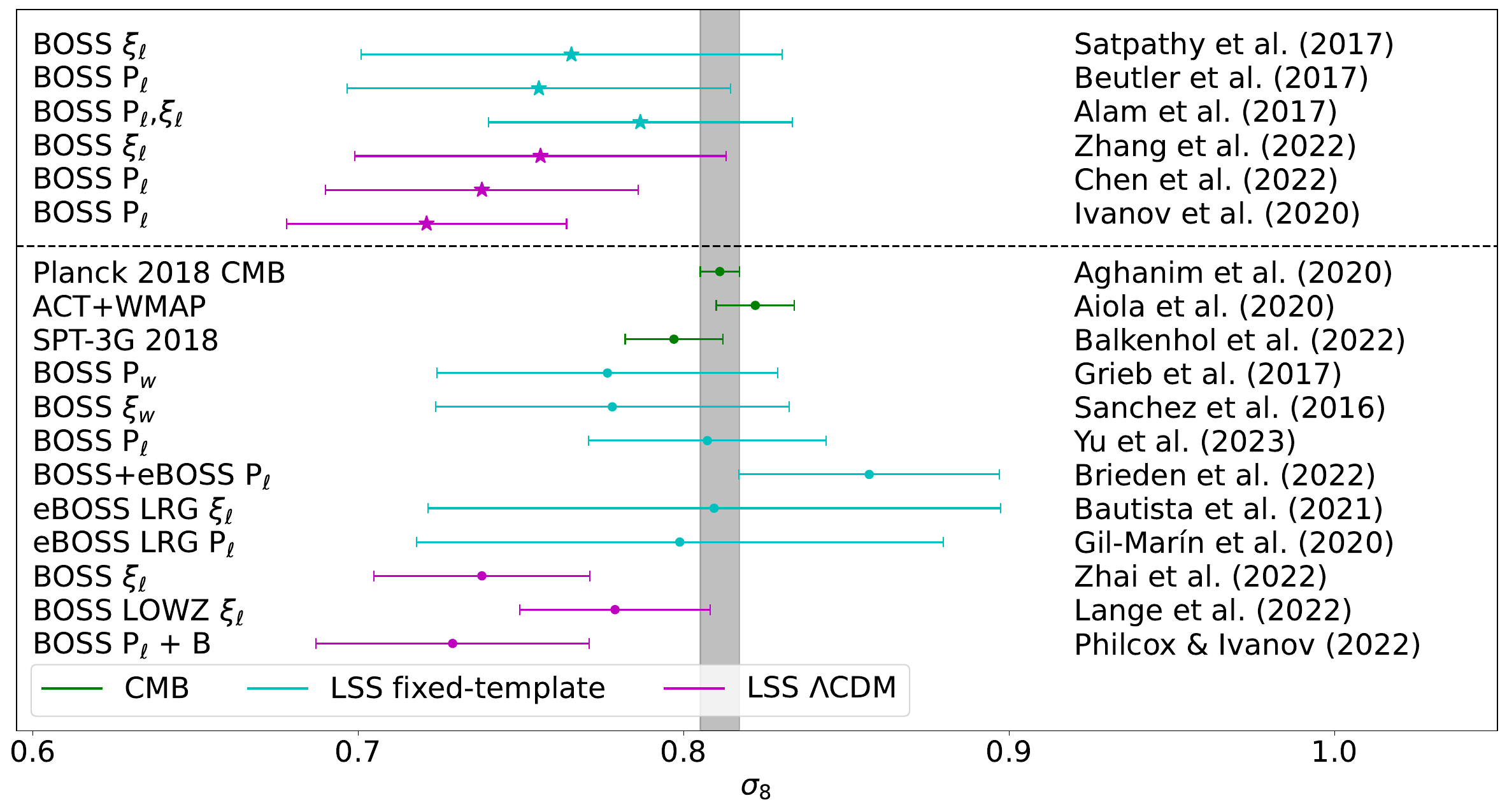}}
    \caption{\change{A representative collection of LSS constraints on $\sigma_8$ from the literature, comparing the fixed template and direct $\Lambda$CDM fitting approaches to the redshift-space distortions of the BOSS DR12 power spectrum and correlation function multipoles. For completeness, we additionally show below the dashed line constraints from RSD fits on extensions to the BOSS DR12 dataset, compared to the inferences from the $\Lambda$CDM model conditioned on the Planck CMB data (see text). The constraints based on CMB analyses are shown in green while the cyan and magenta points come from LSS measurements with fixed-template and $\Lambda$CDM fitting methods, respectively. The starred points refer to the specific dataset and 2-point functions considered in this paper. For the fixed template approaches results are usually reported as constraints on $f\sigma_8$, in which case we divide by the fiducial value of $fD$ in order to convert to $\sigma_8$. We also note that the analyses of Lange et al. (2022) and Zhai et al. (2022) focused on small-scale clustering rather than the large-scale structure growth modeled in this work. \cite{Satpathy17,Beutler17,Alam17,Zhang21b,Chen22,Ivanov20BOSS,PlanckParams18,Aiola_2020,2022arXiv221205642B,Grieb17,Sanchez17,Yu22,Lange2022,Brieden22,Bautista21,Gilmarin20,Philcox22,Zhai2022} }
    }
    \label{fig:sig8_summary}
\end{figure}

Gravitational instability within a (cold) dark matter dominated Universe makes precise predictions for how much the \change{structures} should have grown between the time of the CMB, when the amplitude is measured to be 1 part in $10^4$, until the present day.  How well this prediction matches observations is currently a vexed issue within cosmology \cite{CosmologyIntertwined}. \change{A problem currently facing the LSS community is the apparent discrepancy between different analyses of RSD clustering in the Baryon Oscillation Spectroscopic Survey (BOSS; \cite{Dawson13}), and their comparison} with the expectations of $\Lambda$CDM conditioned upon the Planck data \cite{PlanckLegacy18,PlanckParams18}. Fig.~\ref{fig:sig8_summary} gives an indication of the issue.  Though the significance of any discrepancies is modest, it is of concern that the implied level of agreement with Planck from different analyses of \emph{the same survey} seems so dependent upon methodology.  It is important to understand what is driving the differences in order to ensure that we are ready to analyze the significantly more constraining data sets we expect from the next generation of surveys.  Unfortunately the points in Fig.~\ref{fig:sig8_summary} represent analyses using different summary statistics, data combinations and window functions, different models of non-linearity, bias and RSD as well as different parameters, priors and assumptions.  One difference that separates the earlier BOSS-collaboration analyses (see ref.~\cite{BOSS_DR12} and supporting papers) from the more recent reanalyses \cite{DAmico20,Zhang21b,Kobayashi21,Ivanov:2021zmi,Chen22} is the assumption of a fixed template spectrum in order to compress the data for later reuse vs.\ a direct fit within the $\Lambda$CDM parameter space. \change{To be precise, the BOSS collaboration took several different approaches to their RSD analyses, including both a fixed template approach and direct fits to $\Lambda$CDM.  However for the $\Lambda$CDM fits the Planck CMB data were included which very tightly constrain the shape of $P_{\rm lin}$ such that this is very similar to a template fit.  These different analyses were then reduced to a likelihood involving $f\sigma_8$, $\alpha_\parallel$ and $\alpha_\perp$ and combined with minimum variance weights.  For eBOSS template fits were the \emph{de facto} analysis method.} 

\change{An investigation of the differences between parameter compression vs. the direct fitting methods} while holding the input data, window functions, priors and other modeling choices fixed is the purpose of this paper.  Aspects of this issue have been investigated before (e.g.\ refs.~\cite{Briedan21,Brieden22_2,Ballinger96,Reid16})\change{, but without a detailed comparison between methods while holding the model fixed.} \change{We leave for future work precise comparisons of direct vs template analyses beyond the power spectrum (e.g. bispectrum).}


To understand the methodological differences, recall that within $\Lambda$CDM and its simplest extensions primary CMB anisotropies constrain well the parameters that affect the shape of the matter power spectrum.  Thus the CMB provides a ``calibrated standard spectrum'' that can be used for cosmological inference \cite{PlanckLegacy18}.  For a galaxy survey, then, the remaining degrees of freedom are the conversions from angles and redshifts into (comoving) distance --- which depends upon the late-time expansion history --- and the amplitude of the spectrum after the nearly thousand-fold growth since $z\simeq 1100$.  This motivated the BOSS team to compress the results of their RSD and BAO fits into three numbers describing distances transverse and along the line of sight ($\alpha_\perp$ and $\alpha_\parallel$; see below) and a measure of the amplitude: $f\sigma_8$ \cite{BOSS_DR12}.  A similar compression was performed by eBOSS \cite{eBOSS:2020yzd}.

Since then interest has grown in fitting to the galaxy survey data excluding (most of) the constraints from the CMB, with the BOSS data set being the first that was really large enough to make this approach feasible without significant priors on the model space.  In such an approach the free parameters are those of the underlying model ($\Lambda$CDM in our case) and the predictions of the model are fit directly to the BOSS data \cite{DAmico20,Zhang21b,Kobayashi21,Ivanov:2021zmi,Chen22}.  This allows a test for the consistency of the constraints between the high- and low-redshift Universe.

In this paper we address in some detail whether there is any discrepancy between the ``$\Lambda$CDM'' and ``template'' approaches and if so from what it arises.  We use the same input data and theoretical model in the two approaches. \change{We begin by describing the data employed in Section~\ref{sec:data}. The theory and RSD modeling approaches are discussed in Section~\ref{sec:model}. We describe our analysis and results in Section~\ref{sec:analysis}, where we identify two primary effects causing the discrepancy in $f\sigma_8$, namely a degeneracy with background geometry and prior volume effects. We summarize our conclusions in Section~\ref{sec:conc}.} 


Throughout we shall follow the modeling, parameter and data choices of ref.~\cite{Chen22}, except that for most of our fits we use only the pre-reconstructed power spectrum and do not include the post-reconstruction correlation function multipoles. We use \texttt{CLASS} \cite{CLASS} to compute linear power spectra and background observables including distances. In order to model non-linearity, bias and RSD we use the 1-loop Lagrangian perturbation theory code \texttt{velocileptors} described in refs.~\cite{Chen20,Chen21}. We perform MCMC using \texttt{Cobaya} \cite{CobayaSoftware}. The inclusion of the post-reconstruction correlation function does not qualitatively change our conclusions (see Table \ref{tab:all_res}), but it slightly obscures the cause of the differences.

\section{Data}
\label{sec:data}

\begin{figure}
    \centering
    \resizebox{\textwidth}{!}{\includegraphics{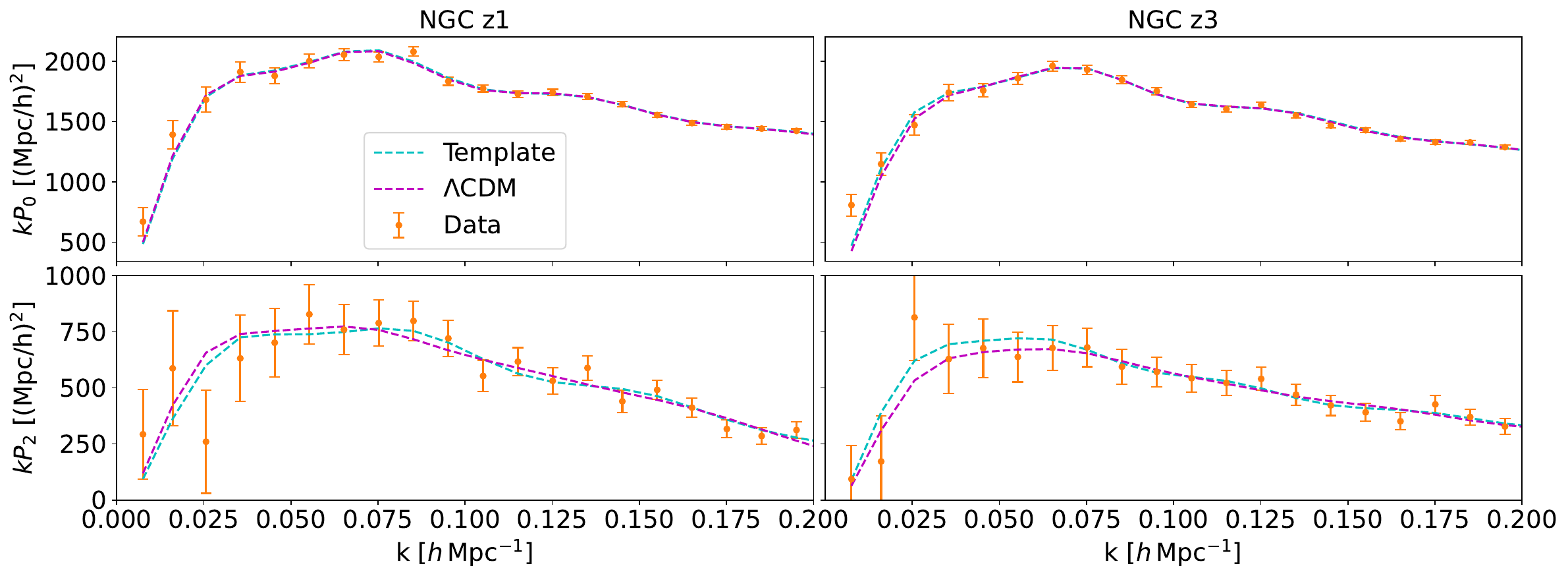}}
    \caption{The BOSS pre-reconstruction power spectrum multipole data for the NGC (orange points) along with the best-fit power spectra from the template (cyan, dashed) and $\Lambda$CDM (magenta, dashed) analyses (see text). The left panels show the \textbf{z1} data, while the panels on the right show \textbf{z3}.
    }
    \label{fig:Pells}
\end{figure}

To illustrate the differences in the context in which they originally arose, we analyze the clustering of galaxies drawn from the final data release of the BOSS galaxy redshift survey \cite{Dawson13}, part of the Sloan Digital Sky Survey III \cite{SDSSIII}.
The final BOSS sample comprises 1,198,006 galaxies in total over 10,252 square degrees of sky.  We use the low and high redshift subsamples, \textbf{z1} and \textbf{z3}, covering $0.2<z<0.5$ and $0.5<z<0.75$ respectively \cite{Reid16,Alam17} and further divide each bin into galaxies observed in the Northern (NGC) and Southern (SGC) \change{Galactic} caps.  Our \change{analysis} builds upon the pipeline developed in ref.~\cite{Chen22}, and we refer the reader to that paper for details.  \change{We make use of the power spectra, and briefly the correlation function multipoles} measured in refs.~\cite{Beutler17,Vargas18,Beutler21} with a theoretical model based upon 1-loop Lagrangian perturbation theory \cite{Chen20,Chen21}.
Fig.~\ref{fig:Pells} shows the power spectrum monopole and quadrupole moments for the NGC data in the \textbf{z1} and \textbf{z3} redshift slices, along with model fits that shall be described later.  The level of agreement is qualitatively similar for the SGC data and for the post-reconstruction correlation function data \cite{Chen22}.

The BOSS two-point function measurements were computed by converting angles and redshifts into distances assuming a flat $\Lambda$CDM cosmology with present-day matter density $\Omega_{M, \rm fid} = 0.31$. This implies that the reported redshift-space power spectrum is related to its value in the coordinates of the true cosmology by
\begin{equation*}
    P^{\rm obs}_s(\bk_{\rm obs}) = \alpha_\perp^{-2} \alpha_\parallel^{-1} P_s(\bk)
    \quad , \quad
    k^{\rm obs}_{\parallel, \perp} =  \alpha_{\parallel, \perp}\  k_{\parallel, \perp}
    \quad ,
\end{equation*}
where the Alcock-Paczynski (AP) parameters\footnote{We do not include a factor of the sound horizon at the drag epoch, $r_{\rm drag}$, in our definition of $\alpha_i$ in contrast to e.g.\ the $\alpha_i$ defined in ref.~\cite{Alam17}.  Our $\alpha_i$ thus isolate the purely `geometric' information.  The equivalent scaling including the $r_{\rm drag}$ factor will be denoted $\tilde{\alpha}$.} are defined as \cite{Alcock79,Padmanabhan08}
\begin{equation}
    \alpha_\parallel = \frac{H^{\rm fid}(z)}{H(z)}
    \quad , \quad
    \alpha_\perp = \frac{D_A(z)}{D^{\rm fid}_A(z)} \quad .
\label{eqn:AP}
\end{equation}
with $D_A$ is the comoving angular diameter distance to redshift $z$.
The equivalent relations for the correlation function are simply the Fourier transforms of the above equations.  We will find it convenient to use a ``dilation'' and ``warping'' parameterization, rather than $\alpha_\parallel$ and $\alpha_\perp$ directly, as in general the isotropic component is much better constrained than the anisotropic with the latter more correlated with the constraints on the amplitude.  Specifically we define \cite{Padmanabhan08}
\begin{equation}
  \alpha = \left( \alpha_\parallel \alpha_\perp^2 \right)^{1/3}
  \quad , \quad
  \epsilon = \left( \alpha_\parallel/\alpha_\perp \right)^{1/3} - 1
  \quad .
\end{equation}
Within $\Lambda$CDM the warp, $\epsilon$, is a function of $\Omega_m$ and is independent of $h$.  It is related to (the inverse of) the parameter $F_{AP}=D_AH/c$ that is sometimes referred to as \textit{the} AP parameter and it causes a mixing of the monopole and quadrupole.  For small $\epsilon$ \cite{Padmanabhan08}
\begin{equation}
  P_0\to P_0 - \frac{2\epsilon}{5}\frac{dP_2}{d\ln k} - \frac{6\epsilon}{5}P_2
  \quad , \quad
  P_2\to \left(1-\frac{6\epsilon}{7}\right)P_2 - \frac{4\epsilon}{7}\frac{dP_2}{d\ln k} - 2\epsilon\frac{dP_0}{d\ln k}
  \quad .
\end{equation}
In our case the two corrections to the monopole approximately cancel, but this is not true \change{for} the quadrupole whose amplitude is modified by changing $\epsilon$.  Since both $\epsilon$ and $f\sigma_8$ modulate the amplitude of $P_2$, we expect them to be correlated.  Note also that a non-zero $\epsilon$ modifies the BAO signal in $P_2$ due to the inclusion of an out-of-phase component ($dP_\ell/d\ln k$) from the monopole and quadrupole.

When we do the template fitting we follow ref.~\cite{Alam17} and (implicitly) scale $\mathbf{k}$ by a further factor of $r_{\rm drag}$, under the assumption that the bulk of the information on the scale comes from the BAO feature and not from the broad-band shape of the power spectrum (for further discussion see ref.~\cite{Chen22}).  The value of $\epsilon$ is clearly independent of $r_{\rm drag}$.  This scaling largely accounts for the cosmology dependence of BAO feature in the template, and can be simply implemented by interpreting our $\alpha$ in the template fit as including an additional $(r_{\rm drag}^{\rm fid}/r_{\rm drag})$, i.e.
\begin{equation}
    \tilde{\alpha} = \alpha \left(\frac{r_{\rm drag}^{\rm fid}}{r_{\rm drag}}\right)
\end{equation}
This aligns our template-fitting procedure with the one most commonly used (e.g.\ refs.~\cite{Alam17,Alam20}).  We also quote $\tilde{\alpha}$ rather than $\alpha$ from the $\Lambda$CDM fits for a more meaningful comparison.

\section{Model and analysis methods}
\label{sec:model}

\subsection{LPT and the Redshift space Power Spectrum}
We model the formation of structures in the universe within a Lagrangian Perturbation Theory (LPT) framework, in which fluid elements are mapped from their initial Lagrangian coordinates, $\bq$, to their observed positions, $\bx$, via the displacement field $\Psi(\bq,\tau)$, such that $\bx=\bq+\Psi(\bq,\tau)$. Nonlinear evolution is handled by expanding the displacements perturbatively ($\Psi =\Psi^{(1)}+\Psi^{(2)}+\Psi^{(3)}...$) in the equation, $\ddot{\Psi}+\mathcal{H}\dot{\Psi}=-\nabla_{\bx}\Phi$, which describes the dynamics of the displacement under Newtonian gravity and expanding spacetime. Since redshift-space distortions are caused by the LOS component of velocities, the transformation to redshift-space in LPT is performed by boosting the displacement field by the velocity $\boldsymbol{u}$ along the LOS direction $\hat{n}$:

\begin{equation}
    \Psi_s = \Psi + \frac{(\hat{n}\cdot\boldsymbol{u})\hat{n}}{\mathcal{H}}.
\end{equation}
In the plane-parallel approximation, we assume that the LOS vector $\hat{n}$ is constant. Number conservation requires that the initial and final galaxy densities are related by $\rho_g(\bx)d^3\bx = \rho_g(\bq)d^3\bq$, which in Fourier space corresponds to
\begin{equation}
    1+\delta_g(\bk) = \int d^3\bq\ e^{i\bk \cdot (\bq + \Psi)} F(\bq).
\end{equation}
The bias functional $F(\bq)$ relates the galaxy density $\rho_g(\bq)$ to the initial conditions $\delta_0$, and can be perturbatively expanded to
\begin{equation}
    F(\bq) = 1 +  b_1\delta_0 + \frac{1}{2}b_2(\delta_0(\bq)^2 - \left\langle\delta_0^2\right\rangle)+b_s(s_0^2(\bq) - \left\langle s^2\right\rangle),
\end{equation}
where $\delta_0$ is the initial Lagrangian overdensity and $s_0$ is the initial shear tensor, given by $s_0= (\partial_i\partial_j/\partial^2 - \delta_{ij}/3)\delta_0$. The power spectrum can now be expressed as 
\begin{equation}
    P_s(\bk) = \int d^3\bq \left\langle e^{i\bk \cdot (\bq + \Delta_s)}F(\bq_1)F(\bq_2) \right\rangle_{\bq = \bq_1-\bq_2},
    \label{eq: P_int}
\end{equation}
where $\Delta_s = \Psi_s(\bq_1) - \Psi_s(\bq_2)$. An important aspect of the LPT model we use in this work is the resummation of long-wavelength displacements, which is necessary to properly model the damping of the BAO peak. To do so, the linear part of the pairwise displacement is split into long- and short-wavelength contributions separated by an infared scale $k_{\rm IR}$, with the former kept resummed in the exponent using the cumulant theorem.

In addition to the above, effective-theory contributions must be included to control the sensitivity of our PT model to small-scale physics, so that the complete power spectrum in our parametrization is given by
\begin{equation}
    P_s(\bk) = P_s^{PT}(\bk) + (\alpha_0 + \alpha_2\mu^2)P_{\rm Zel}(\bk) + (\text{SN}_0 + \text{SN}_2\bk^2\mu^2),
\end{equation}
including counterterms $\alpha_{0,2}$ and stochastic contributions, $\text{SN}_{0,2}$. Here $P_{\rm Zel}$ refers to the linear matter power spectrum with infrared linear displacements resummed.

The theory code that we employ, \texttt{velocileptors}, computes the 1-loop LPT predictions for the power spectrum multipoles, including the bias, RSD, counter and stochastic terms listed above and infrared (IR) resummation as described earlier. A more detailed description of this code and validation with N-body simulations can be found in refs.~\cite{Chen20,Chen21}, which also contain references to the broader literature detailing the various physical effects described above. In the analysis presented in this paper, we refer to $\{(1+b_1)\sigma_8$, $b_2$, $b_s$, $\alpha_0$, $\alpha_2$, SN${}_0$, and SN${}_2\}$ as "nuisance" parameters which are allowed to differ between NGC and SGC subsamples in our joint fits. Table~\ref{tab:priors} lists the priors applied to these parameters.

\begin{table}[t!]
\label{tab:priors}      
\centering                          
\begin{tabular}{c|c|c|c}        
$\Lambda$CDM & Template & Bias & Stoch/Counter \\ \hline \hline
H$_0$ &  $f\sigma_8$ & $(1+b_1)\sigma_8$ & $\alpha_0$ \\
$\mathcal{U}[60,80]$ & $\mathcal{U}[0,2]$ & $\mathcal{U}[0.5,3.0]$ & $\mathcal{N}[0,100]$ \\
\hline
$\Omega_{m}$ &  $\alpha_{\parallel}$ & $b_2$ & $\alpha_2$ \\
$\mathcal{U}[0.2,0.4]$ & $\mathcal{U}[0.5,1.5]$ & $\mathcal{N}[0,10]$ & $\mathcal{N}[0,100]$ \\
\hline
$\log(10^{10} A_\mathrm{s})$ &  $\alpha_{\perp}$ & $b_s$ & SN${}_0$ \\
$\mathcal{U}[1.61,3.91]$ & $\mathcal{U}[0.5,1.5]$ & $\mathcal{N}[0,5]$ & $\mathcal{N}[0,1000]$ \\
\hline
&  & & SN${}_2$ \\
&  & & $\mathcal{N}[0,50000]$ \\
\end{tabular}
\caption{Priors on parameters used in the $\Lambda$CDM and template fitting methods. The $\Lambda$CDM model involves H$_0$, $\Omega_{m}$, $\log(10^{10} A_\mathrm{s})$ and all of the bias, stochastic, and counterterms. The template method fits $f\sigma_8$, $\alpha_{\parallel}$ and $\alpha_{\perp}$ as well as the same bias, stochastic and counterterms.  The entries  $\mathcal{U}[{\rm min,max}]$ and $\mathcal{N}[\mu,\sigma]$ refer to uniform and Gaussian normal distributions, respectively.}    
\end{table}

\subsection{Fitting approaches}
We concern ourselves with two approaches to fitting the RSD 2-point functions, which we refer to hereafter as the ``template'' and ``$\Lambda$CDM'' methods. In the template approach, a fiducial $\Lambda$CDM cosmology is chosen that determines the shape of the linear power spectrum. This template power spectrum is kept fixed while the observed data is compressed into three parameters to be varied (in addition to the bias, stochastic, and counterterms); namely the two distance scalings, $\alpha_{\parallel}$ and $\alpha_{\parallel}$, and the amplitude given by the product $f\sigma_8$. More specifically, the growth rate of structures, $f$, controls the monopole-to-quadrupole ratio and is approximately given by $f \simeq \Omega_m^{0.55}$. $\sigma_8$ is the total amplitude of the power spectrum at 8 $h^{-1}$Mpc scales. The ``standard'' template procedure is to hold $\sigma_8$ fixed to $\sigma_8^{\rm fid}$ while varying $f$, and then interpreting the result as $f\sigma_8$. This is the method that we follow when discussing template fits in this paper. An alternative approach would be to vary both $f$ and $\sigma_8$ and reporting the product as $f\sigma_8$; these two methods can lead to a difference of up to $0.5\,\sigma$ in $f\sigma_8$ constraints. However, when letting $\sigma_8$ vary in addition to $f\sigma_8$, regions of parameter space with very low $\sigma_8$ are explored --- if we demand a reasonable $\sigma_8$ (such as a lower bound as generous as $\sigma_8 > 0.5$) we get the same result as the standard method. That being said, the low values of $\sigma_8$ (implying unphysically high $f>1$ and by extension $\Omega_m>1$) could be an issue with the template method. This will be explored further and reported on in future papers. 

The alternate method of fitting RSD correlation functions or power spectra involves a direct fit of the parameters underlying the cosmological model, in our case $\Lambda$CDM. In this method the shape of the linear power spectrum is free to vary, and the ratio of high-to-low $k$ amplitude depends on $\Omega_m$ and $h$. In general, the $\Lambda$CDM parameters are $\{h$, $\Omega_{cdm}$, $\Omega_b$, $A_s$, $n_s$, $M_\nu\}$; however, $n_s$, $M_\nu$, and $\Omega_b$ are much more tightly constrained by Planck and/or BBN than is currently achievable with LSS observations, and thus we restrict our parameter space to only $h$, $A_s$ and $\Omega_m$, while fixing ($n_s, M_\nu, \Omega_b = 0.9665, 0.06, 0.02242$). The parameters and priors of the $\Lambda$CDM and template fit methods are also listed in Table~\ref{tab:priors}. 

One of the differences between the two methods is that the template fit is model-independent whereas in direct fitting approaches a specific cosmological model (e.g.\ $\Lambda$CDM) is chosen that determines the parameters being varied. The advantage of the template fit in this regard is that the compressed parameters ($f\sigma_8,\alpha_{\parallel},\alpha_{\perp}$) can be interpreted within the parameter space of any model of choice (satisfying the assumptions). In addition, the compressed parameters of the template fit only depend on the late-time geometry and dynamics of the universe while the dependence on the physics of earlier epochs that enters the transfer function is fixed by the chosen template. The $\Lambda$CDM method does not separate the dependence of the early-time physics through the shape of the transfer function and late-time geometry, and the additional information from the shape of the tranfer function results in tighter constraints on parameters such as $\Omega_m$ and $h$ that also affect the geometrical properties of the late-time universe. In this sense, the model-independence of the template method sacrifices some of its constraining power when compared to the $\Lambda$CDM fit. 

The improvement in constraints on cosmological parameters in the $\Lambda$CDM method comes at a cost. This approach requires a Boltzmann code such as \texttt{CLASS} or \texttt{CAMB} to compute the transfer function for the linear power spectrum at every step of an MCMC, making the fit more computationally expensive compared to the template method. However, we can alleviate this problem by using an emulator to approximate both the transfer function and the LPT computations from \texttt{velocileptors} for any given set of parameters. Our emulator is based on a Taylor series in the model parameters centered around a reasonable set of values, ($\Omega_m, h, \sigma_8 = 0.31, 0.68, 0.739$). We refer readers to Appendix A of ref.~\cite{Chen22} for more details about this Taylor series emulator and its accuracy. We train a similar emulator for the template method, centered around reference values of ($f\sigma_8,\alpha_{\parallel},\alpha_{\perp} = 0.46, 1.0, 1.0$). As a result, the template and $\Lambda$CDM fits converge in approximately equal times.

Finally, an extension to the standard template fit has been proposed that attempts to preserve the extra information captured by the $\Lambda$CDM method while still being a model-independent fit through compressed physical parameters \cite{Briedan21}. This ``ShapeFit'' method involves a modification to the linear power spectrum via the ansatz
\begin{equation}
    P'_{\rm lin}(\bk) = P_{\rm lin}(\bk)\ \exp\left\{ \frac{m}{a}\tanh \left[a\ln\left(\frac{k}{k_p}\right) \right] + n\ln\left(\frac{k}{k_p}\right) \right\}, 
\end{equation}
that depends on a scale-dependent slope parameter, $m$, and the scale-independent slope, $n$. The pivot scale, $k_p\approx \pi/r_d^{\rm fid} \approx 0.03 h^{-1}$Mpc is the scale at which the slope of the hyperbolic tangent is maximum. Since the scale-independent slope, $n$, is fully degenerate with the spectral tilt, $n_s$, and we fix $n_s$ in our fits, in practice we set $n=0$ and only vary $m$. The ShapeFit method does improve on the constraining power achieved by the template fit; however, it suffers from the same issue mentioned earlier of exploring unphysical regions of parameter space when fitting $f$ and $\sigma_8$ independently. For the remainder of this paper we restrict our attention to the standard template and direct fitting $\Lambda$CDM methods, but a more detailed comparison that includes the ShapeFit method will be left for future work using simulated data. Finally, we note that the three methods agree in their constraints when Planck priors are applied to early-time physics. The differences we investigate in this work appear in the limit that LSS analyses are sufficiently powerful to provide independent cosmological constraints, with only a prior on $\omega_b$ from BBN.

\section{Analysis}
\label{sec:analysis}

We perform two classes of fits to the BOSS data, using the same model for bias, non-linearity and redshift-space distortions and the same input data vector but differing in the parameters being varied.  The first is a ``template'' fit in which we hold the linear theory power spectrum shape fixed to that of a $\Lambda$CDM cosmology with ``fiducial'' parameters and vary only the amplitude and the AP parameters.  In the second we vary the linear power spectrum shape by sampling $\Omega_m$ and $h$ in addition to the power spectrum amplitude.  The AP parameters are self-consistently computed from the value of $\Omega_m$ assuming the $\Lambda$CDM expansion history.


\begin{figure}
    \centering
    \resizebox{\textwidth}{!}{\includegraphics{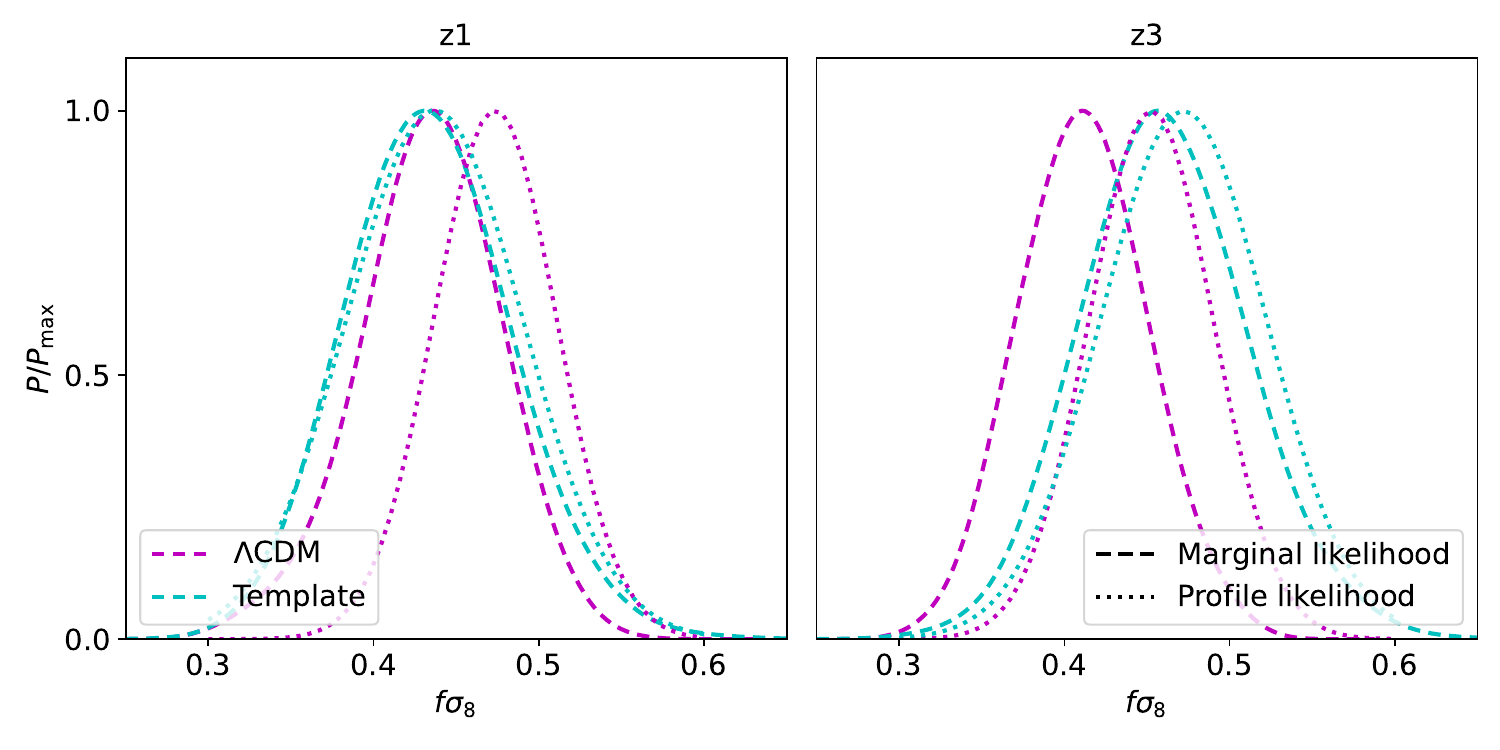}}
    \caption{Marginalized posteriors (dashed) and profile likelihoods (dotted) of $f\sigma_8$ for the \textbf{z1} (left) and \textbf{z3} (right) redshift bins of BOSS.  All of the curves have been normalized to peak at unity for ease of comparison. The results from the template fit are shown in cyan and the $\Lambda$CDM results are shown in magenta.  While the template constraints are in good agreement between the two methods (cyan dashed and dotted lines), the $\Lambda$CDM results show an offset with the marginal likelihood (magenta dashed line) peaking at lower $f\sigma_8$ than the profile likelihood (magenta dotted line).}
\label{fig:z1z3_fs8}
\end{figure}

First we shall investigate for which of the subsamples of the BOSS data the differences between the two approaches is \change{the} largest. Table~\ref{tab:all_res} shows the marginalized $f\sigma_8$ constraints from each of the two approaches for the low- and high-redshift bin, fitting either the NGC, SGC, or both \change{Galactic} cap subsamples together. In addition, Fig.~\ref{fig:z1z3_fs8} shows the marginalized $f\sigma_8$ posteriors from each of the two approaches for the two redshift bins. When fitting the NGC and SGC data jointly we see that the two analysis methods give very similar results for marginal posteriors of the \textbf{z1} bin ($f\sigma_8 = 0.434 \pm 0.051$ and $0.434 \pm 0.044$ for the template  and $\Lambda$CDM respectively), but for \textbf{z3} the fixed template approach prefers a noticeably higher $f\sigma_8$ ($0.461 \pm 0.051$) than the direct fit ($0.414 \pm 0.040$).  The preference for higher $f\sigma_8$ in the template approach holds for both the NGC and the SGC data in \textbf{z3}. For the combined NGC and SGC sub-samples we fit a single set of $f\sigma_8$ and AP parameters, but fit two independent sets of bias parameters for the NGC and SGC data points. Going forward we perform tests on the joint \textbf{z3} data to try to isolate the cause of the difference in $f\sigma_8$, but show only the curves with the NGC bias parameters and data when appropriate.

Table \ref{tab:all_res} shows results including and excluding the post-reconstruction BAO data.  Henceforth we will focus on the pre-reconstruction power spectrum multipoles.  The inclusion of the post-reconstruction, correlation function (BAO) data shifts the $f\sigma_8$ distributions to higher values, for both the template and $\Lambda$CDM fits, but does not significantly alter the sense or size of the difference between the two methods (Table \ref{tab:all_res}).  We will omit the post-reconstruction BAO data from our fits in order to highlight the impact of choosing a template vs.\ a cosmological model when fitting the pre-reconstruction power spectrum multipoles. 

Figure \ref{fig:z1z3_fs8} highlights two features that we wish to address.  One is the shift between the marginal posteriors for the template and $\Lambda$CDM fits.  The second is the shift to lower $f\sigma_8$ of the marginal posterior compared to the profile likelihood (dotted line in Fig.~\ref{fig:z1z3_fs8}) for the $\Lambda$CDM model, with a shift in the same direction but of much smaller amplitude in the case of the template fit. \change{The profile likelihood is a method for parameter estimation when the likelihood $\mathcal{L}(\theta,\bf\lambda)$ depends on both a parameter of interest ($\theta$) and nuisance parameters ($\bf\lambda$) that `profiles' out the nuisance parameters by maximizing the likelihood $\mathcal{L_{\theta}(\bf\lambda)}$ at fixed $\theta$. This is different from the marginal likelihood which integrates over $\bf\lambda$ instead of maximizing them. In our case, $\theta = f\sigma_8$ and $\bf\lambda$ refer to the remaining ($\alpha_{\parallel}$,$\alpha_{\perp}$, bias, stochastic, and counterterms) parameters. For Gaussian posteriors the profile and marginal likelihoods are expected to agree, but for non-Gaussian posteriors prior volume effects in the marginalization can cause them to differ, making the profile likelihood a useful diagnostic for when these effects occur. The observed discrepancy between $f\sigma_8$ from the two likelihood distributions suggests that some parameters are too poorly constrained for the posteriors to be Gaussian. This discrepancy is more significant for the $\Lambda$CDM than the template fit, suggesting} that prior volume effects and a degeneracy with the power spectrum shape are causing a shift of the peak of the marginal posterior away from the best-fitting model.  We shall take each of these two features in turn.

\begin{table}
\centering
\begin{tabular}{ccccc}
  & Galactic & Pre/Post & \multicolumn{2}{c}{$f\sigma_8$} \\
 $z_{\rm eff}$  &  Cap  & Recon & $\Lambda$CDM (best-fit) & Template (best-fit) \\ [0.5ex] 
 \hline
 \hline
 \multirow{4}{*}{0.38} & SGC & Pre &  $0.399\pm 0.072$ (0.475) & $0.494^{+0.110}_{-0.130}$ (0.501)\\[0.5ex] 
   & NGC & Pre & $0.457\pm 0.047$ (0.479)& $0.429\pm 0.059$ (0.420)\\[0.5ex] \cline{2-5}  
   & \multirow{2}{*}{Joint} & Pre & $0.434^{+0.045}_{-0.040}$ (0.476)& $\vphantom{\Big[}0.434^{+0.047}_{-0.053}$ (0.435)\\ [0.5ex]
   &  & Post & $0.425\pm 0.037$ (0.434) &  $0.430\pm 0.045$ (0.419)\\[0.5ex]
 \hline \hline
 \multirow{4}{*}{0.59}  & SGC & Pre & $0.369\pm 0.061$ (0.436) & $\vphantom{\Big[}0.385^{+0.087}_{-0.100}$ (0.372)\\[0.5ex]
   & NGC & Pre& $0.434\pm 0.048$ (0.461)& $0.492^{+0.057}_{-0.069}$ (0.483)\\[0.5ex] \cline{2-5}
   & \multirow{2}{*}{Joint} & Pre & $0.414\pm 0.040$ (0.445) & $0.461\pm 0.051$ (0.485)\\[0.5ex]
   &  & Post & $0.429\pm 0.037$ (0.442) & $0.483^{+0.042}_{-0.047}$ (0.486)\\[0.5ex]
   \hline \hline
\end{tabular}
\caption{Marginalized $f\sigma_8$ constraints from $\Lambda$CDM and template analyses (pre and post BAO reconstruction) of the different redshift bin and galactic cap subsamples of BOSS.  In each case we quote the mean and $1\,\sigma$ error from the marginalized posterior, along with the value of the best-fit model of the chains in parentheses. 
}
\label{tab:all_res}
\end{table}

\subsection{Degeneracy with background geometry}

First we investigated whether the discrepancy seen in Fig.~\ref{fig:z1z3_fs8} was caused by an unfortunate choice of template.  We switched the template from that chosen by the BOSS team\footnote{The BOSS fiducial cosmology has $\Omega_m=0.31$, $h=0.676$, $\Omega_bh^2=0.022$, $\sigma_8=0.8$ and $n_s=0.97$ \cite{Alam17}. The sound horizon scale is $r_d = 147.3\,$Mpc. For this cosmology $fD=0.5831$ at $z=0.38$ and $0.5756$ at $z=0.61$, which allows an easy conversion from $f\sigma_8(z)$ to $\sigma_8$.} to that given by the best-fit $\Lambda$CDM model in our alternative analysis.\footnote{The template from our best-fit $\Lambda$CDM model has $\Omega_m=0.318$, $h=0.699$, $\sigma_8=0.767$, with $\Omega_bh^2$ and $n_s=0.97$ fixed to the BOSS fiducial values. For this cosmology $r_d = 143.9\,$Mpc.}  Fixing $P_{\rm lin}$ to this best fit we find that $\epsilon$ and $f\sigma_8$ are virtually unchanged ($f\sigma_8$, $\epsilon = 0.461 \pm 0.051$, $0.992^{+0.044}_{-0.052}$ for the original $P_{\rm lin}$ and $0.472 \pm 0.050$, $0.984^{+0.042}_{-0.048}$ for the fit using the best-fit $\Lambda$CDM model), while $\tilde\alpha$ shifts to keep the BAO peak fixed in physical coordinates ($0.995 \pm 0.015$ to $1.011 \pm 0.015$). The $\simeq 0.2\,\sigma$ shift in $f\sigma_8$ seen here, when compared to the $\simeq 1\,\sigma$ shift between template and $\Lambda$CDM methods, highlights the robustness of the template method to modest changes in the template but indicates that the discrepancy observed in Fig.~\ref{fig:z1z3_fs8} must arise due to the fixing of $P_{\rm lin}$ rather than the specific shape of $P_{\rm lin}$. 

\begin{figure}
    \centering
    \resizebox{\textwidth}{!}{\includegraphics{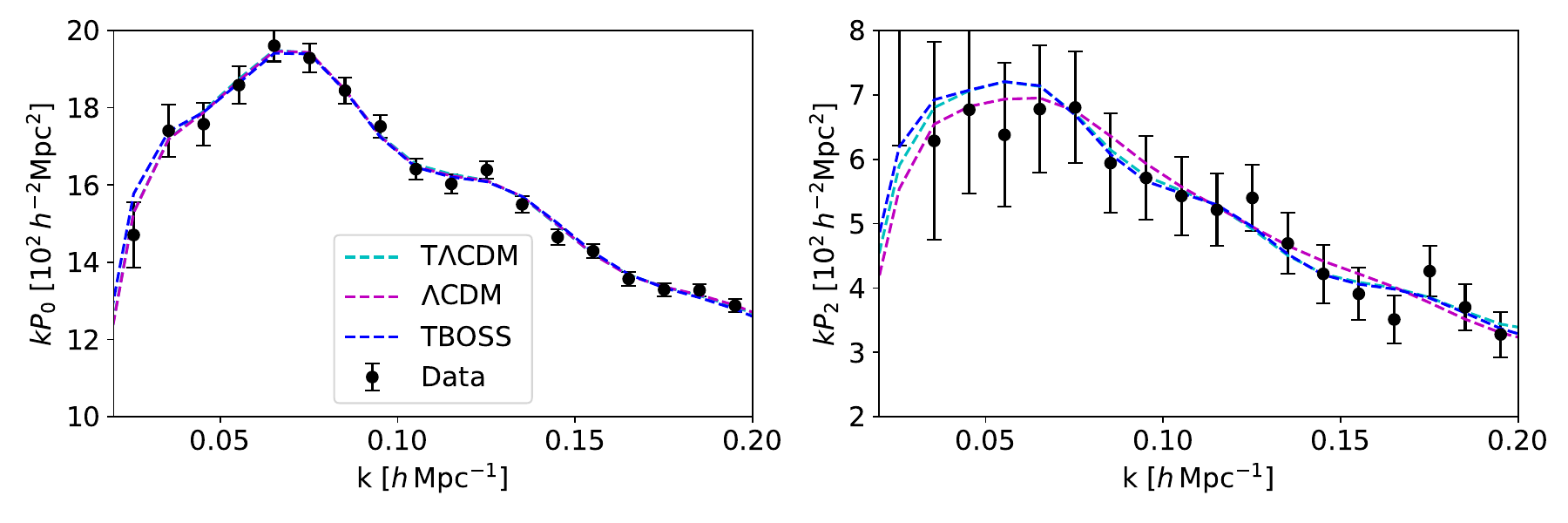}}
    \caption{Best-fit model $P_\ell(k)$ fit to NGC\textbf{z3} data (points with errors) for the template fit using $\Lambda$CDM best-fit as the fixed template (cyan), using the fiducial BOSS cosmology as the template (blue) and the $\Lambda$CDM fit (magenta).
    (Left) The monopole, $P_0$ and (Right) the quadrupole, $P_2$. In the monopole plots, the cyan T$\Lambda$CDM curve is underneath the magenta $\Lambda$CDM curve. 
    }
\label{fig:TLCDM_pell}
\end{figure}

In Table~\ref{tab:chi2} we show the monopole, quadrupole, and total $\chi^2$ values for the two template fits with different, fixed $P_{\rm lin}$ as well as for the $\Lambda$CDM fit. We refer to the template fit that uses $P_{\rm lin}$ from the best-fitting \change{$\Lambda$CDM model as ``T$\Lambda$CDM''} and the template fit with the original BOSS template $P_{\rm lin}$ as ``TBOSS''. When choosing the $\Lambda$CDM template instead of the BOSS one we see an overall improvement in the fit. When compared to $\Lambda$CDM, the monopole terms fit about as well; however there is a noticeable improvement in the quadrupole. If we take the best fit ($f\sigma_8$, $\tilde\alpha$, $\epsilon$) values and map them to $\Lambda$CDM parameters, we find that T$\Lambda$CDM prefers a cosmology with ($\Omega_m$, $h$, $\log(10^{10} A_\mathrm{s})$)=(0.443, 0.652, 2.661), whereas the best-fit result from $\Lambda$CDM (which, recall, was used as the fiducial template for T$\Lambda$CDM) was ($\Omega_M$, $h$, $\log(10^{10} A_\mathrm{s})$) = (0.318, 0.699, 2.793).
So even when using the $\Lambda$CDM best-fit as the fiducial template, the template fit is shifting $\tilde{\alpha}$ and $\epsilon$ in a way that corresponds to drastically different $\Omega_m$ and $h$.

We show in Fig.~\ref{fig:TLCDM_pell} the model $P_\ell$(k) curves from the $\Lambda$CDM and T$\Lambda$CDM fits. Consistent with the $\chi^2$ values, we do not observe a significant difference in the behaviors of the monopole terms for the two models. In the quadrupole, however, the T$\Lambda$CDM curve appears to better fit the data points at $k\gtrsim0.075\ihmpc$ that oscillate above and below the best-fit $\Lambda$CDM curve, with worse agreement for $k<0.075\ihmpc$. The $\Lambda$CDM curve passes in an almost straight line through the high-$k$ data points, whereas T$\Lambda$CDM can shift $\epsilon$ in a way that introduces residual oscillations that better pass through the quadrupole data points. The consequence of this is a poorer agreement with the data at lower $k$, but due to the larger error bars in this regime there is little $\chi^2$ penalty. We also computed the correlation function quadrupoles for these models.  While these oscillatory differences are more pronounced in configuration space they were nonetheless relatively minor (and well within the errorbars of the data), consistent with the power spectrum analysis.

\begin{table}
\centering
\begin{tabular}{c|ccc|ccc|cccc}
 & \multicolumn{6}{c|}{$\chi^2$} & \\
 & \multicolumn{3}{c}{NGC} & \multicolumn{3}{c}{SGC} & \multicolumn{3}{|c}{best-fit} \\
 Type  &  $P_0$  & $P_2$ & Total & $P_0$  & $P_2$ & Total & $f\sigma_8$ & $\tilde{\alpha}$ & $\epsilon$\\ [0.5ex] 
 \hline 
 $\Lambda$CDM  & 11.5 & 9.3 & 21.0 & 15.5 & 13.3 & 28.7 & 0.445  & 1.020 & -0.001\\ 
 TBOSS         & 14.8 & 7.4 & 22.5 & 13.6 & 13.7 & 27.4 & 0.485 & 0.993 & -0.017\\
 T$\Lambda$CDM & 11.7 & 8.3 & 19.9 & 15.3 & 13.6 & 29.0 & 0.470 & 1.002 & -0.018
\end{tabular}
\caption{Minimum $\chi^2$ values for the $\Lambda$CDM fit and the two template fits with different, but fixed, $P_{\rm lin}$ (see text).  The $\chi^2$ values correspond to fits to NGC\textbf{z3} data and are broken up into the contribution from the monopole, the quadrupole and both $P_0$ and $P_2$ (including their correlation).  The full fit has 26 degrees of freedom so all of the best fits are statistically acceptable.
}
\label{tab:chi2}
\end{table}

\begin{figure}
    \centering
    \resizebox{\textwidth}{!}{\includegraphics{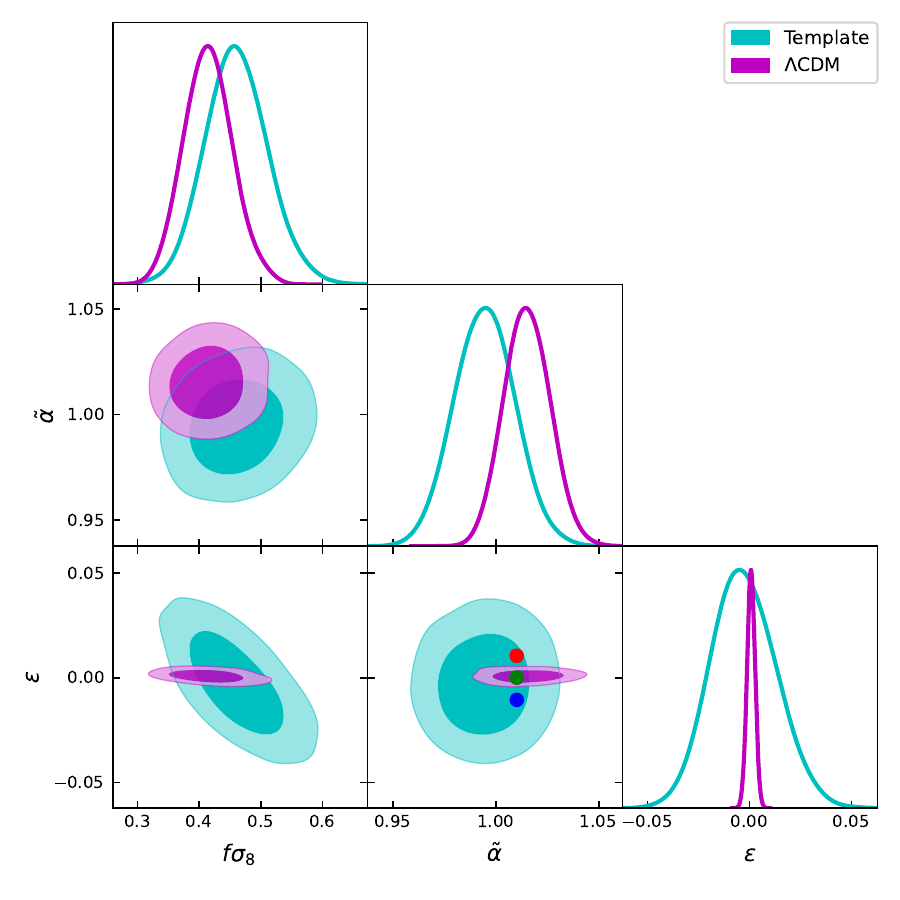}}
    \caption{Marginalized posterior distributions for the key cosmological parameters from the template (cyan) and $\Lambda$CDM (magenta) fits to the BOSS Joint \textbf{z3} pre-reconstruction power spectrum multipoles.
    Three example models, discussed further in the text, are shown as the red, green and blue dots in the $\tilde{\alpha}-\epsilon$ panel.    }
\label{fig:z3_temp_fs_iso}
\end{figure}

It is well known \cite{Ballinger96,Reid12} that in the template-based approach $f\sigma_8$ is correlated with $\epsilon$, since both affect the amplitude of the quadrupole (see \S\ref{sec:data}).  We see this in the lower left panel of Fig.~\ref{fig:z3_temp_fs_iso}.  By contrast the $\Lambda$CDM fits explore a much narrower range of $\epsilon$ and no such degeneracy is apparent (we shall return to the reason for this below).  In particular this means that a higher value of $f\sigma_8$ can be compensated by a lower value of $\epsilon$ with little $\chi^2$ penalty in the template approach, and since $\epsilon$ is constrained less well than $\alpha$ this has the ability to affect the preferred $f\sigma_8$.  We checked that all of the values of $\alpha$ and $\epsilon$ explored by the template chain in Fig.~\ref{fig:z3_temp_fs_iso} can be realized within $\Lambda$CDM for some value of $\Omega_m$ and $h$.  Indeed this is the case, however some of the values correspond to cosmologies that are not preferred by the BOSS data themselves due to the implied change in the shape of $P(k)$.

To further demonstrate that the change in $f\sigma_8$ is being driven by $\epsilon$, we perform a fit with $f\sigma_8$ \change{along with all biases and nuisance terms free}, while $\tilde\alpha$ and $\epsilon$ are held fixed at the values derived from the best-fit cosmology of the $\Lambda$CDM analysis. We find that when switching from the $\tilde\alpha$ and $\epsilon$ preferred by the template to those preferred by $\Lambda$CDM, $f\sigma_8$ shifts from $0.461\pm 0.051$ to $f\sigma_8=0.433\pm 0.035$, consistent with the shift in marginalized constraint seen in the two analyses.

\begin{figure}
    \centering
    \resizebox{\textwidth}{!}{\includegraphics{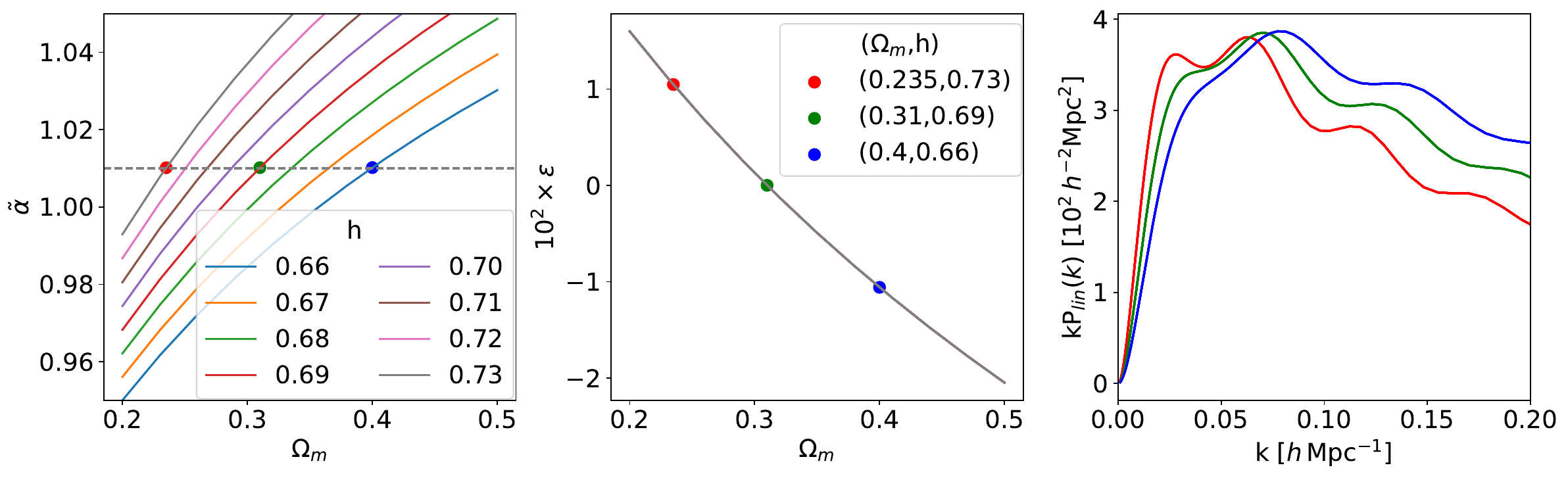}}
    \caption{\textit{Left}: $\tilde\alpha$ at $z=0.59$ as a function of $\Omega_m$ for different values of $h$. Markers indicate example cosmologies for which $\alpha$ remains constant that we shall further explore. \textit{Middle}: $\epsilon$ at $z=0.59$ as a function of $\Omega_m$ (it is independent of $h$).
    \textit{Right}: $k\,P_{\rm lin}(k)$ for the three example cosmologies from the previous panels.
    Note that the $\tilde\alpha$ and $\epsilon$ values for these cosmologies all lie well within the $1\,\sigma$ contour of the template fit (Fig.~\ref{fig:z3_temp_fs_iso}).
    }
    \label{fig:fixed_aiso_models}
\end{figure}

Figure \ref{fig:fixed_aiso_models} shows how $\tilde\alpha$ and $\epsilon$ at $z=0.59$ depend upon $\Omega_m$ and $h$ within $\Lambda$CDM.  Note that the range of $\tilde\alpha$ and $\epsilon$ explored by the template chain covers a broad range of $\Lambda$CDM models, particularly in $\Omega_m$.  Changes in $\Omega_m$ of this magnitude have a large effect on the shape of the linear theory power spectrum, $P_{\rm lin}$, predicted by $\Lambda$CDM as can be seen in the right panel of Fig.~\ref{fig:fixed_aiso_models}.  Such large changes in shape turn out to be highly disfavored by the BOSS data, which is why the $\Lambda$CDM chain does not explore a wide range of $\epsilon$ values.  This in turn means that the higher values of $f\sigma_8$ that are preferred by the template fit, corresponding to lower values of $\epsilon$, are ruled out in the $\Lambda$CDM chain by the shape of the power spectra.

In Fig.~\ref{fig:fixed_aiso_models} we have chosen three sets of ($\Omega_m$, $h$) values for which $\tilde{\alpha}$ is approximately fixed but $\epsilon$ varies.  The tight constraint on $\tilde{\alpha}$ fundamentally arises because of the well-detected BAO feature in the high S/N monopole measurement. The value of $\Omega_m$ represented by the green marker is the same as the fiducial template cosmology, as well as being within $1\,\sigma$ of the posterior mean value of the $\Lambda$CDM fit. The other $\Omega_m$ values are more than $4\,\sigma$ away from what is preferred by the $\Lambda$CDM fit. We see in the third panel \change{that there} is much variety in the shape of the linear power spectrum for each of these sets of cosmological parameters. The template fit is not penalized by this because the linear power spectrum is fixed, so it has the freedom to choose AP parameters as needed to fit the data.  Due to the $\epsilon-f\sigma_8$ degeneracy the value of $f\sigma_8$ can shift accordingly. This freedom is not available to the $\Lambda$CDM analysis.  For $\Lambda$CDM there is only a narrow range of $\epsilon$ that can fit the data because of how drastically the implied change in $\Omega_m$ would affect the linear power spectrum shape.

\begin{figure}
    \centering
    \resizebox{\textwidth}{!}{\includegraphics{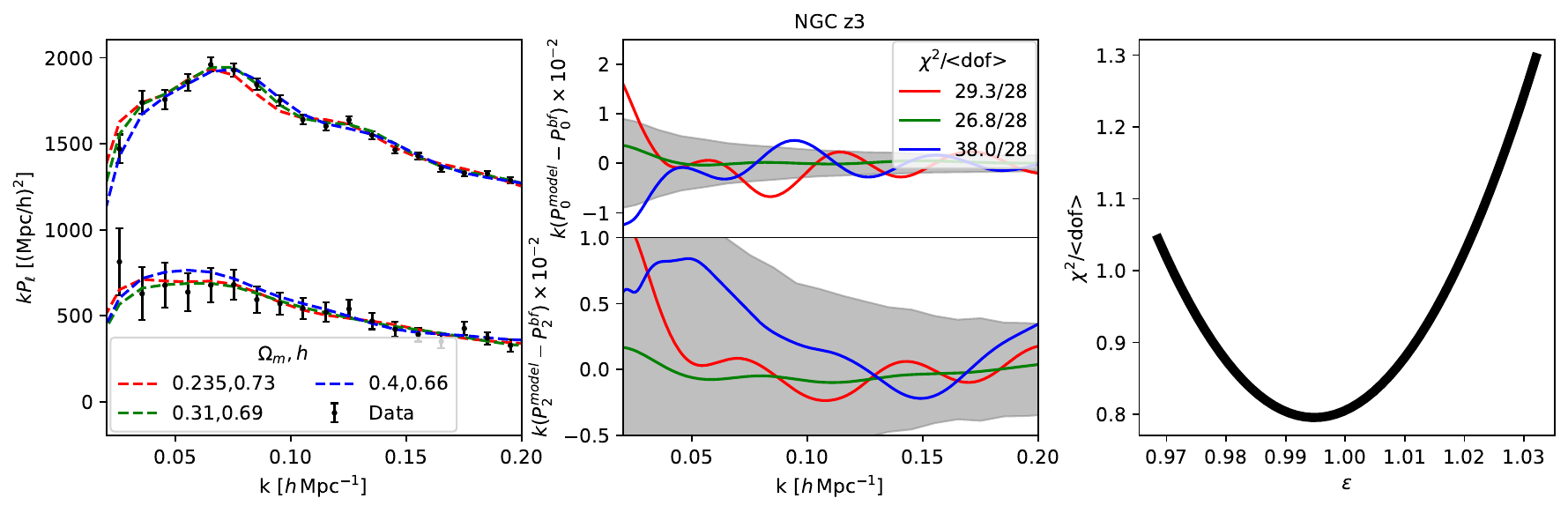}}
    \caption{\textit{Left}: $k\,P_{\ell}(k)$ curves for the three different pairs of ($\Omega_m$, $h$) values of Fig.~\ref{fig:fixed_aiso_models}, where the bias parameters and $\log(10^{10} A_\mathrm{s})$ are adjusted to best fit the NGC \textbf{z3} data (points with error bars). \textit{Middle}: Difference between the $P_{\ell}$ models for the three cosmology pairs and the $P_{\ell}$ of the best-fit $\Lambda$CDM model. The upper and lower panels correspond to the monopole and quadrupole components respectively. The shaded bands show the $1\,\sigma$ errors of the NGC \textbf{z3} data. \textit{Right}: Reduced $\chi^2$ as a function of $\epsilon$, from fitting a quadratic function to a series of points with varying ($\Omega_m$, $h$). }
\label{fig:fixed_aiso_Pells}
\end{figure}

Figure \ref{fig:fixed_aiso_Pells} gives further details.  For each of our three, example cosmologies (red, green and blue points in Fig.~\ref{fig:fixed_aiso_models}) the left panel shows the predicted multipoles (after accounting for the BOSS NGC window function) compared to the data.  The central panel shows the residuals to the best-fit model, with the $\chi^2$ values marked while the right panel shows how the $\chi^2$ depends upon $\epsilon$ for the $\Lambda$CDM model.  Note how the $\Lambda$CDM fits highly disfavor the values of $\epsilon$ that are preferred by the template model, due to the implied change in $P_{\rm lin}(k)$ shape associated with the $\Omega_m$ and $h$ values which map to such an $\epsilon$ in $\Lambda$CDM. It is worth noting that models with more freedom in the expansion history than $\Lambda$CDM could allow for more flexibility of this kind \cite{Ivanov20BOSS}.

\begin{figure}
    \centering
    \resizebox{\textwidth}{!}{\includegraphics{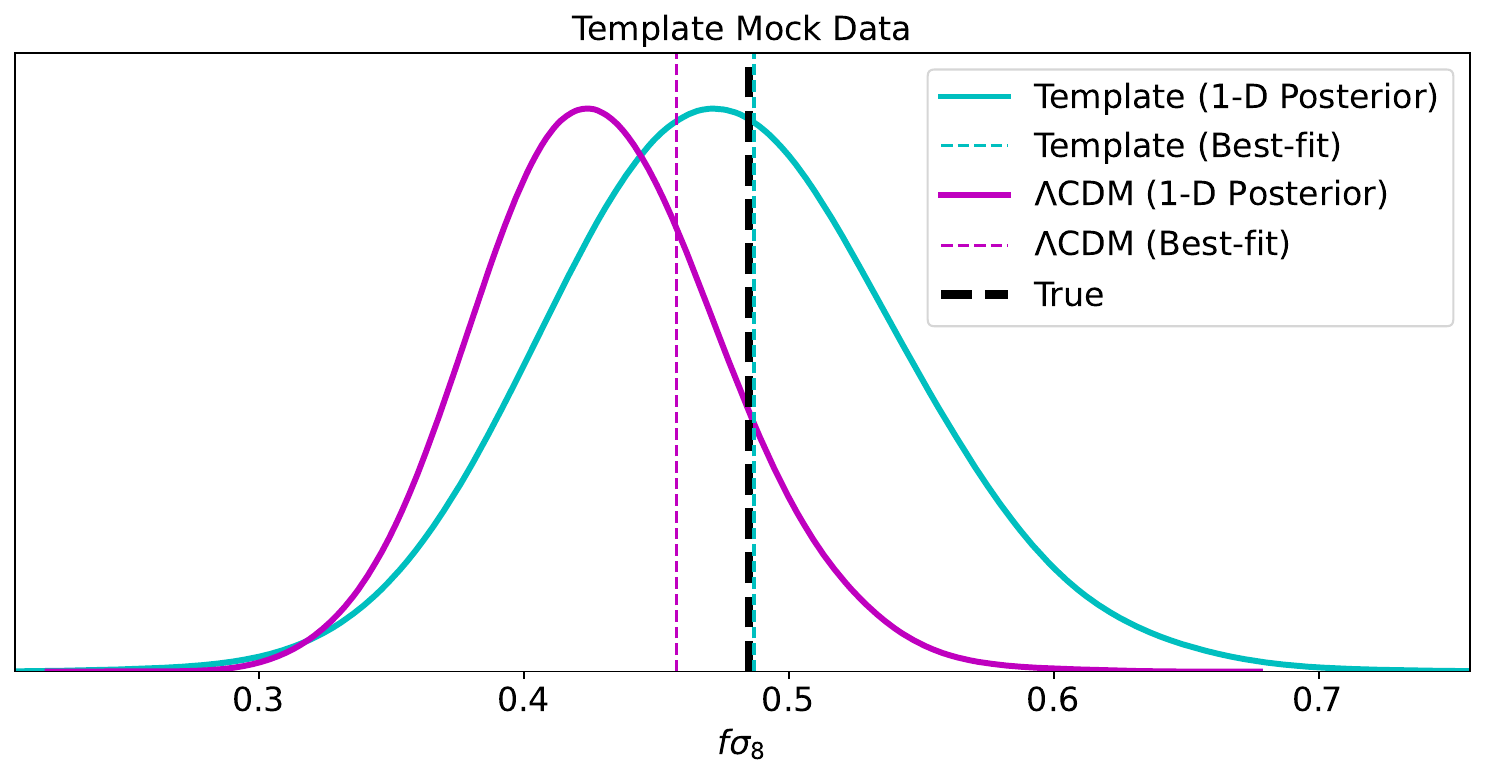}}
    \caption{Template and $\Lambda$CDM fits to mock data generated from the best-fitting template model.  While the template fit returns unbiased constraints the $\Lambda$CDM fit does not recover the correct $f\sigma_8$ because the required $\epsilon$ and shape of $P_{\rm lin}$ are not simultaneously allowed within $\Lambda$CDM. 
    }
    \label{fig:mock_temp}
\end{figure}

If $\Lambda$CDM is the correct model, then the preference for ``low'' $\epsilon$ in the \textbf{z3} data must be due to a noise fluctuation that affects the template and $\Lambda$CDM fits quite differently.  In the case of the template fit, a change in the shape of the $P_\ell$ to better match the data can be obtained by varying $\epsilon$.  It would be up to external data (e.g.\ a Planck prior) to disfavor these points.  For the $\Lambda$CDM fit the additional constraint that the AP parameters be consistent with the shape of $P_{\rm lin}$ within the $\Lambda$CDM paradigm `regularizes' this behavior so that such values of $\epsilon$ are not explored.

As a cross-check of this hypothesis, we create mock $P_\ell(k)$ data using the $\Lambda$CDM theory model at the best fit parameters for the \textbf{z3} sample. We use the same covariance matrix and window functions as is used for the real BOSS data and repeat both the $\Lambda$CDM and template fits.  Since the data are noiseless we expect be able to recover the correct model parameters and indeed both the template and $\Lambda$CDM fits give consistent $f\sigma_8$ posterior distributions with best-fit values almost exactly matching the true $f\sigma_8$.  As a final test, we generate noiseless data from the best-fit template model and repeat the exercise.  While the template fit returns unbiased constraints the $\Lambda$CDM fit is not able to recover the correct $f\sigma_8$, reinforcing the idea that the combination of background cosmology and $P_{\rm lin}(k)$ inferred from the template mock data lies outside the allowed space for $\Lambda$CDM.
In Fig.~\ref{fig:mock_temp} we show the $f\sigma_8$ posteriors and best fit results for the template and $\Lambda$CDM fits of the mock data created from the template best-fit parameters.

\subsection{Nuisance parameter selection and priors}

Now we turn to the second effect mentioned in \S\ref{sec:analysis} and highlighted in Fig.~\ref{fig:z1z3_fs8}, the shift between the marginal posterior and the profile likelihood.  Fig.~\ref{fig:z1z3_fs8} shows that both the profile likelihood and the marginalized posterior agree (for both \textbf{z1} and \textbf{z3}) when performing a template fit.  However for the $\Lambda$CDM model the marginalized posterior peaks $\approx 1\sigma$ below the profile likelihood (see \cite{Chen22} for similar discussion).  For \textbf{z1} the marginalized posterior agrees better with the template approach while for \textbf{z3} it is the profile likelihood that shows the best agreement.  In both cases, parameter volume effects that arise due to degeneracies within the model cause the posterior peak to be shifted with respect to the maximum likelihood point, and we wish to investigate this in more detail.

\begin{figure}
    \centering
    \resizebox{\textwidth}{!}{\includegraphics{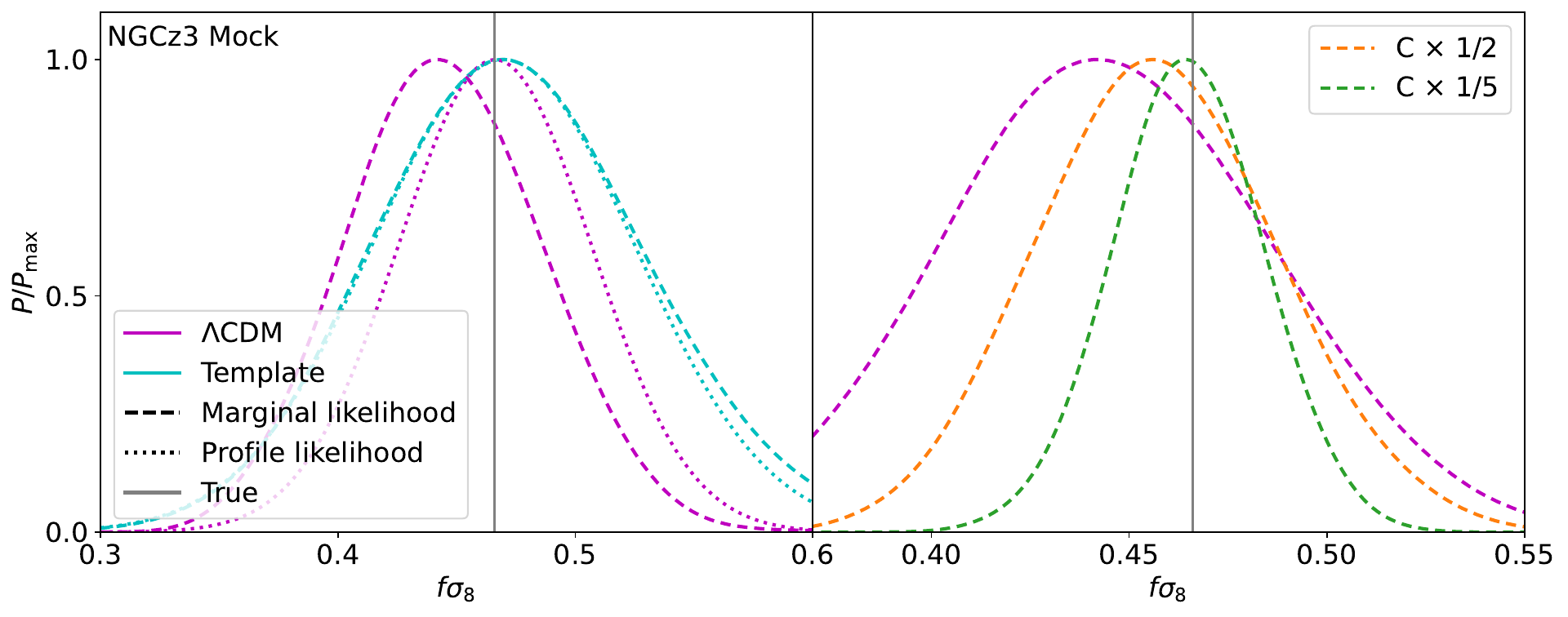}}
    \caption{\textit{Left}: Template and $\Lambda$CDM marginal (dashed) and profile (dotted) likelihoods from fitting to mock data generated with the fiducial BOSS template parameters. While the marginal and profile likelihoods agree with the true value of $f\sigma_8$, there is a 0.5$\sigma$ shift between the marginal and profile likelihood constraints peaks of the $\Lambda$CDM fit. \textit{Right}: $\Lambda$CDM marginal (dashed) and profile (dotted) likelihoods as in the left panel, along with marginal posteriors with covariance rescaled by 1/2 (orange) and 1/5 (green).
    }
    \label{fig:mock_fid}
\end{figure}

In order to investigate degeneracies that are present within the $\Lambda$CDM model, we return to mock data with cosmology fixed to the BOSS fiducial cosmology that \change{we use}for our template fits. The model parameters \change{are} simply taken from the best-fit $\Lambda$CDM model of the NGC\textbf{z3} sample, and \change{this is} used to generate ``mock'' data.  The window function and covariance matrix are unchanged. In the left panel of Fig.~\ref{fig:mock_fid} we show the marginal and profile likelihood results of the two methods applied to this mock data. Consistent with the fits on the real data of Fig.~\ref{fig:z1z3_fs8}, we again see that the marginal and profile likelihoods of the template fit agree with one another as well as with the $\Lambda$CDM profile likelihood, all peaking near the true $f\sigma_8$ value of 0.466. The $\Lambda$CDM marginal likelihood is again shifted to a lower $f\sigma_8 = 0.445 \pm 0.041$, and it is this effect which we wish to understand. 

\begin{figure}
    \centering
    \resizebox{\textwidth}{!}{\includegraphics{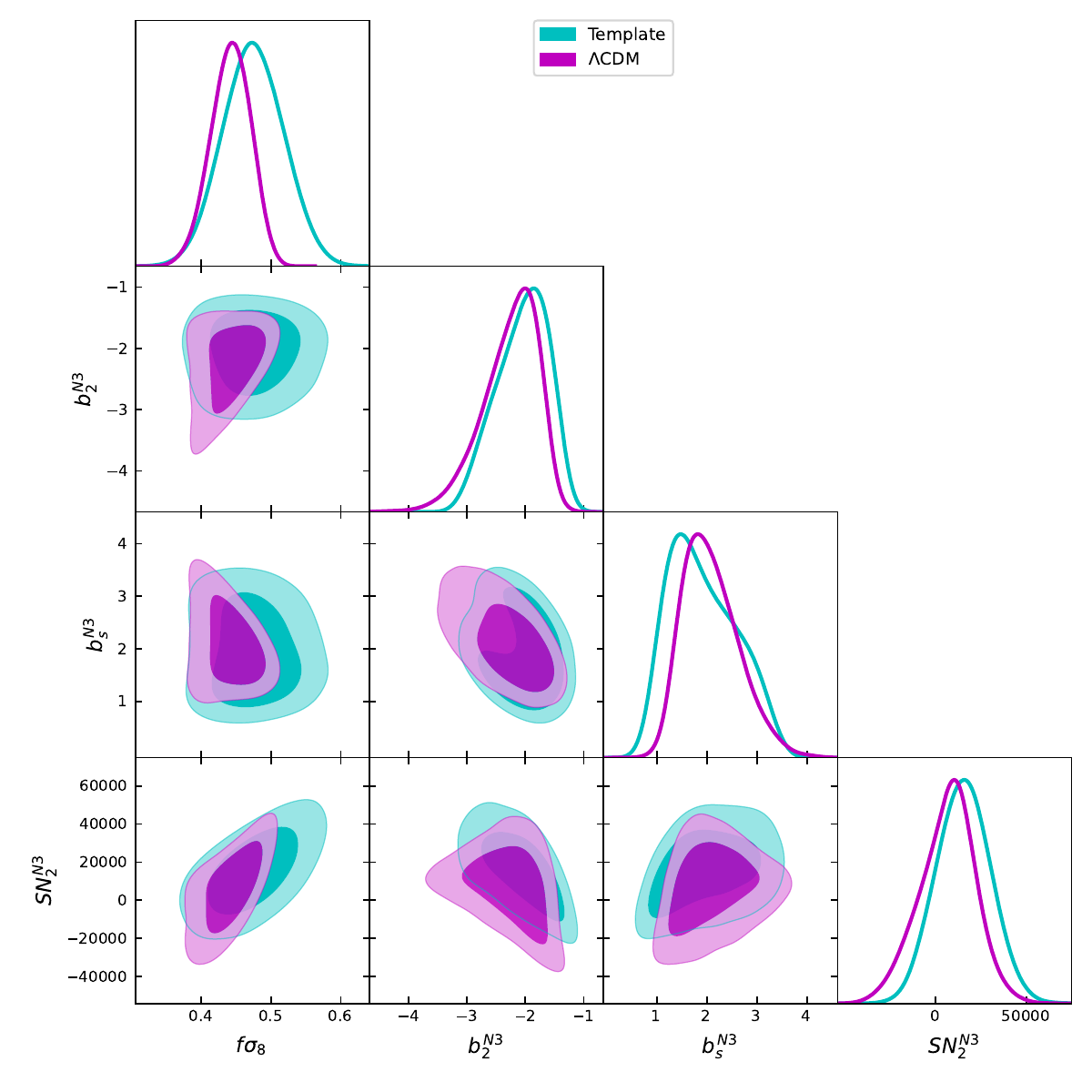}}
    \caption{Marginalized posterior distributions for $f\sigma_8$,$b_2$, $b_s$, and $SN_2$ from the template (cyan) and $\Lambda$CDM (magenta) fits to the NGCz3 Mock data.
    }
    \label{fig:bias_contours}
\end{figure}


Since all of the true parameter values of the mock data are known, we begin by fixing all of the galaxy bias parameters ($(1+b_1)\sigma_8$, $b_2$, $b_s$, $\alpha_0$, $\alpha_2$, SN${}_0$, and SN${}_2$) to their true values\footnote{These values are: $(1+b_1)\sigma_8$ = 1.77, $b_2$ = -1.98, $b_s$ = 1.90, $\alpha_0$ = 1.42, $\alpha_2$ = -1.73, SN${}_0$ = -197 $h^{-3}$ Mpc$^3$., and SN${}_2$ = 13600 $h^{-5}$ Mpc$^5$.} and run a $\Lambda$CDM fit with only $\Omega_m$, $H_0$ and $\log(10^{10} A_\mathrm{s})$ as free parameters.  In this case the $\Lambda$CDM fit almost exactly recovers the correct cosmological constraints ($f\sigma_8 = 0.4667 \pm 0.0061$). This suggests that the downward shift of the marginal likelihood must be caused by a degeneracy between the $\Lambda$CDM and the nuisance parameters. In Table~\ref{tab:priors} we list the priors on all cosmology and bias parameters used in the $\Lambda$CDM and template fitting methods, as a reference to aid in the following discussion.


By fitting $\Lambda$CDM models to the mock data with different bias parameters fixed to their true values we \change{identify} those that are most responsible for the shift in the $f\sigma_8$ constraint, namely $b_2$, $b_s$ and SN${}_2$.  Next, we \change{run} $\Lambda$CDM and template fits with all other parameters held fixed to their ``true'' values keeping free $\Omega_m$, $H_0$, $\log(10^{10} A_\mathrm{s})$, $b_2$, $b_s$ and SN${}_2$ for the $\Lambda$CDM fits and $f\sigma_8$, $\alpha_{\parallel}$, $\alpha_{\perp}$, $b_2$, $b_s$ and SN${}_2$ for the template fits. Fig.~\ref{fig:bias_contours} shows the posterior distributions for $f\sigma_8$, $b_2$, $b_s$ and SN${}_2$ for the two cases. The $\Lambda$CDM constraint is $f\sigma_8=0.442^{+0.031}_{-0.027}$, which is about $0.8\,\sigma$ below the true value.  The template fit has $f\sigma_8 = 0.475 \pm 0.044$ which is within $0.25\,\sigma$ of the truth. 

\begin{figure}
    \centering
    \resizebox{\textwidth}{!}{\includegraphics{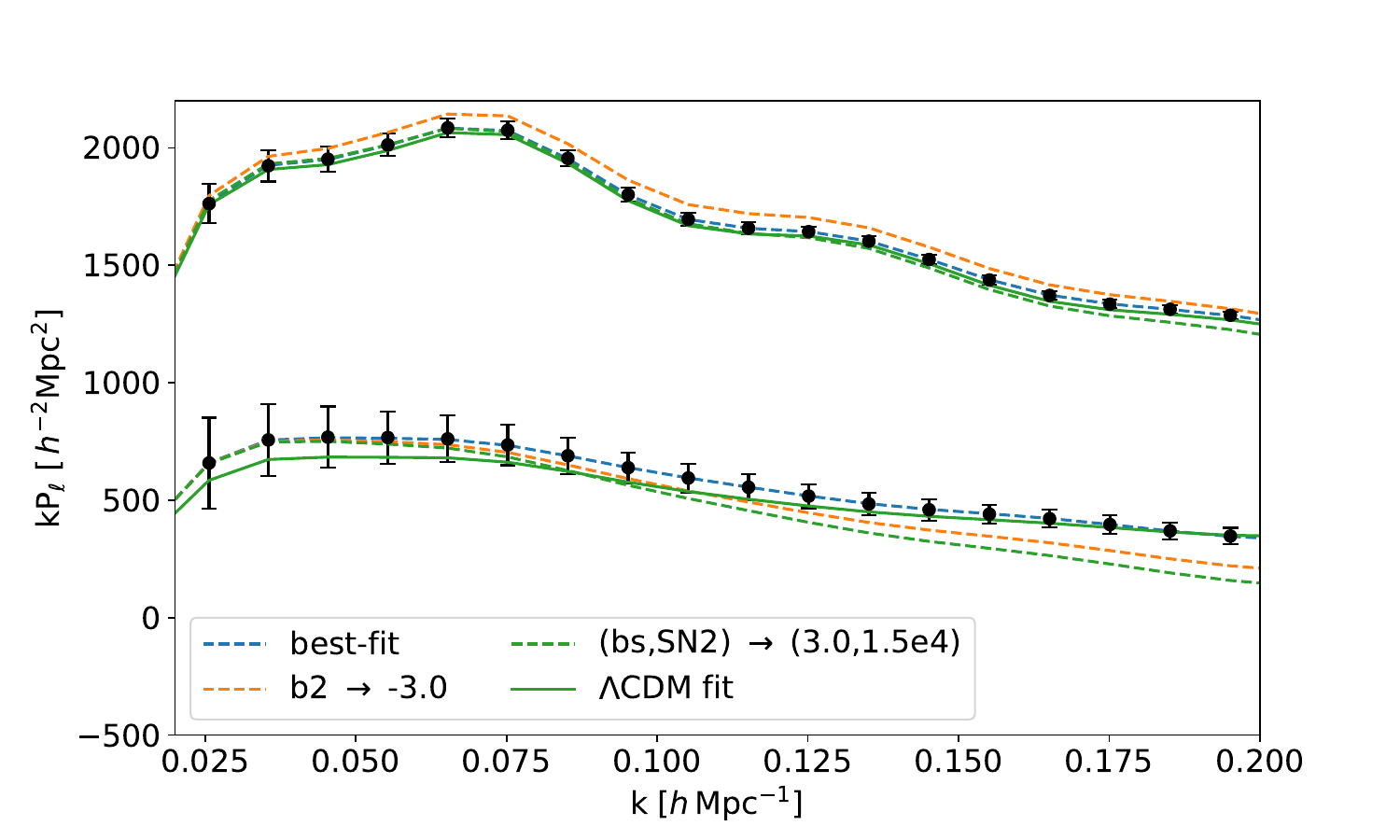}}
    \caption{$P_{\ell}(k)$ curves with different bias parameters fixed. The blue dashed curve shows the best-fit model to the mock data. The orange dashed curve shows the model in which $b_2$ is fixed to -3.0 while the rest of the parameters are kept at the best-fit values. The green curve takes the model of the orange curve but additionally changes $b_s$ and $SN_2$ to 3.0 and 15000 respectively (while $b_2 = -3.0$ still). Finally, the solid green curve has the same bias parameters as the green dashed curve, i.e. ($b_2$, $b_s$, $SN_2$) = (-3.0, 3.0, 15000), but the $\Lambda$CDM parameters ($\Omega_M$, $H_0$, $\log(10^{10} A_\mathrm{s})$) are shifted to best fit the data. These new $\Lambda$CDM values are ($\Omega_M$, $H_0$, $\log(10^{10} A_\mathrm{s})$) = (0.3049, 67.45, 2.813), with $f\sigma_8 = 0.4070$.
    }
    \label{fig:bias_degen_pells}
\end{figure}

We can try to understand the role that the nuisance terms play in this shift in the $\Lambda$CDM constraint by observing the 2D contours between $f\sigma_8$ and the three parameters in the first column of Fig.\ref{fig:bias_contours}. In each case the $\Lambda$CDM contours have a triangular shape that is wider in the vertical direction at lower $f\sigma_8$ and narrows at higher $f\sigma_8$. This implies that a $\Lambda$CDM chain has fewer models to explore at larger $f\sigma_8$ without large $\chi^2$ penalties when compared to the availability of models with acceptable $\chi^2$ in the lower $f\sigma_8$ regime. As a result the marginalized constraint is shifted to lower $f\sigma_8$.  What is quite interesting is that we do not observe this in the template fit. This implies that due to the fixed $P_{\rm lin}(k)$ shape in the template method it is not able to increase the range of acceptable models with extreme values in the nuisance parameters by going into a different $f\sigma_8$ regime. This is because the preferred $f\sigma_8$ is not a byproduct of shape changes in $P_{\rm lin}(k)$ like in the $\Lambda$CDM case. Therefore, while there are degeneracies between $f\sigma_8$ and the nuisance parameters, there is little shift in the marginal constraints. 


\change{In order to further explore} the degeneracy between $P_{\rm lin}$ shape (i.e.\ cosmological parameters) and nuisance parameters that gives rise to the downward shift in the marginal posterior of $f\sigma_8$ we investigated how different parameters affected $P_\ell$.  In Fig.~\ref{fig:bias_degen_pells} we attempt to illustrate the degeneracies \change{between the three} parameters of interest and the shape of the power spectrum. We begin with the best-fitting model of the mock data (blue dashed curve). Taking this model but setting $b_2 = -3.0$ causes the amplitude of $P_0$ to shift up and $P_2(k \geq 0.075)$ to shift down (orange curve). Fig.~\ref{fig:bias_contours} shows a negative correlation between $b_2$ and both $b_s$ and SN${}_2$, suggesting that the effect of lowering $b_2$ to $-3.0$ can be partially mediated by shifting $b_s$ and SN${}_2$ up (to $3.0$ and $15,000$ respectively in our example).  This is shown in the green dashed curve in Fig.~\ref{fig:bias_degen_pells}, where we observe that the amplitude of $P_0$ decreases again to better fit the data, but the high-$k$ part of $P_2$ is shifted down even further.  To show that this can be alleviated by a change in the cosmology we fix these nuisance parameters at ($b_2$, $b_s$, SN${}_2$) = $(-3.0, 3.0, 15000)$, and run a fit with only $\Omega_M$, $H_0$ and $\log(10^{10} A_\mathrm{s})$ free.  The best fit is shown as the green solid curve.  We find that the shape change in the linear power spectrum allowed by $\Lambda$CDM resolves the poor agreement with the data in the high-$k$ part of the $P_2$ curve. Even though the model shifts away from the data in the low-$k$ regime of $P_2$, the error bars are much larger in that regime, so the $\chi^2$ penalty from a poorer fit at low-$k$ is smaller than the improvement in $\chi^2$ achieved by better fitting the high-$k$ points. The $f\sigma_8$ of this final curve is $0.407$ whereas the dashed curves all have $f\sigma_8 = 0.464$, in closer agreement with the true value ($0.466$). 

We note that since the size of the error bars in the low-$k$ part of $P_2$ is what allows these shape changes in the power spectrum to occur without large penalties in $\chi^2$, in the limit of noiseless data this parameter volume effect should disappear. To check this, we ran the full $\Lambda$CDM fits on the mock data after rescaling the covariance by factors of 1/2 and 1/5. These artificial rescalings can be interpreted as if the ``data'' is being measured from regions of the sky with 2$\times$ or 5$\times$ the volume compared to the original BOSS NGC\textbf{z3} sample. The marginal likelihoods using the rescaled covariance are shown in orange and green dashed lines in the right panel of Fig.~\ref{fig:mock_fid}. We find that in the $C\times 1/2$ case the $f\sigma_8$ marginal constraint shifts up from $f\sigma_8 = 0.445 \pm 0.041$ to $f\sigma_8 = 0.457 \pm 0.031$; and for $C\times 1/5$ we get $f\sigma_8 = 0.465 \pm 0.019$, which is in almost exact agreement with the profile likelihood and the ``true'' value of $f\sigma_8$. As expected, the parameter volume effects we see in the $\Lambda$CDM approach become less of an issue as the data become more constraining. \change{For reference, the volume of the BOSS survey is approximately 5 $h^{-3}$Gpc$^3$ \cite{Dawson13} while the forecast for DESI predicts a survey volume of about 50 $h^{-3}$Gpc$^3$ for redshifts between $0.65 < z < 1.65$ \cite{DESI}, which suggests that these prior volume effects will be very small for the final DESI data release.} 

Finally, we \change{check} to what extent these parameter volume effects exist in a \change{ShapeFit} \cite{Briedan21} analysis of the same mock data\change{, which is described in \S\ref{sec:model}}. We find that the shapefit method gives $f\sigma_8 = 0.472 \pm 0.061$, in agreement with the standard template fit. This indicates that the shapefit functional form is sufficiently rigid that the nuisance terms don't interact with the cosmology through degeneracies as they do in $\Lambda$CDM. \change{This result is consistent with ref.~\cite{Brieden22} who also did not find prior volume effects in their ShapeFit analyses of BOSS data. Further testing of ShapeFit, along with comparisons to standard compression and direct fitting methods will be performed on mocks simulating the heightened levels of precision of future LSS surveys.}

\section{Conclusion}
\label{sec:conc}

One of the most puzzling discrepancies in cosmology today is the apparent mismatch between the amplitude of clustering seen at low redshift compared to that predicted by evolution of the CMB fluctuations in a (cold) dark matter dominated Universe.  One of the key observables for determining the amplitude of structure in the local Universe is the anisotropic clustering of galaxies induced by RSD.  Current surveys can put $\mathcal{O}(5\%)$ constraints on the amplitude of clustering, $\sigma_8$, from the study of RSD and we anticipate that future surveys should be able to reduce this uncertainty significantly \cite{DESI,Euclid,Schlegel22}.

Currently, however, different techniques for analyzing galaxy clustering in the BOSS survey come to different conclusions about the amplitude of structure.  While the discrepancy is of only modest statistical significance, the fact that the difference arises entirely from analysis methodology makes it worth understanding the source in some detail. \change{In this paper we take an in-depth examination of two prevailing fitting approaches in the literature. The first, ``template'' method involves fixing the linear power spectrum via a template cosmology and varying the compressed amplitude and distance scaling parameters that can then (in principle) be interpreted in the context of a given cosmological model. The second approach involves directly varying the underlying parameters of a cosmological model ($\Lambda$CDM in our case) to fit the data. The included sensitivity to early-time physics, via the shape of the transfer function, results in tighter constraints on cosmological parameters. Thus far, both fitting approaches have been used in analyses of the BOSS and eBOSS data sets; however, the additional differences in theoretical model, parametrization, window function, and priors obscures the sources of discrepancies between the constraints obtained by these groups.}

In this paper we have investigated the difference between the template fitting approach as compared to a direct, $\Lambda$CDM fit holding the input data and theoretical model fixed. We have found that the former tends to produce higher $f\sigma_8$ values than the latter and the difference arises through a combination of two effects. \change{While differences between the template and direct fitting approaches have been investigated before \cite{Ivanov20BOSS,Briedan21,Brieden22_2}, these studies have tended to compound the differences due to differing \textit{cosmological} degrees of freedom between the two approaches by additionally introducing different \textit{dynamical} models for galaxy clustering, using effective theory-based approaches in the case of $\Lambda$CDM fits but reverting to older models combining perturbation theory with empirical parametrizations for fingers-of-god for template fits. In this paper we have opted to use only the former, which represents the state-of-the-art of our understanding of perturbation theory, in order to perform a clean comparison between the two approaches. By using the same dynamical model, including identical priors on galaxy bias and EFT parameters in both approaches, we are able to tease out two main physical effects at the root of the different $\sigma_8$ constraints between the methods.}

The first effect is the well-known degeneracy between the ``warping'' parameter, $\epsilon$, and the amplitude $f\sigma_8$.  The warping, $\epsilon$, is highly constrained in direct, $\Lambda$CDM fits to the BOSS data because large changes in $\epsilon$ require similarly large changes in $\Omega_m$ that in turn affect the shape of the power spectrum.  Such changes are disfavored by the shape of the power spectrum measured in BOSS reducing the impact of this degeneracy.  In the case of a template fit, however, a broad range of $\epsilon$ is allowed because the power spectrum shape is not self-consistently changed.  This then leads to a broad range of allowed $f\sigma_8$, including higher values than preferred by the $\Lambda$CDM model.
As long as only cosmologies consistent with the template shape are explored, as originally envisioned, the conditional $f\sigma_8$ constraints are consistent between the two techniques.  Indeed, if we put tight priors on the $\Lambda$CDM parameters other than $\sigma_8$ and also put tight priors on $(\tilde{\alpha},\epsilon)$ in the template fit then the results become identical.  However if the full range of $\epsilon$ allowed by the BOSS data alone is explored then an upward shift in $f\sigma_8$ appears that we postulate is due to a noise fluctuation in the \textbf{z3} data that prefers $\epsilon$ inconsistent with the assumed template shape within $\Lambda$CDM.

The second effect arises because perturbative models of galaxy clustering come with a large set of nuisance parameters that can be partially degenerate with cosmological parameters influencing the shape of the linear theory power spectrum (e.g.\ $\Omega_m$ and $h$).  This degeneracy can cause a ``prior volume effect'' in which the peak of the marginalized $f\sigma_8$ posterior is offset from the best-fit model, i.e.\ the peak in the profile likelihood.  In our case such a parameter volume effect results in a shift in the marginal posterior for $f\sigma_8$ approximately $1\,\sigma$ down from the peak of the profile likelihood. To better understand this effect, we worked with mock data mimicking the BOSS NGC\textbf{z3} sample, and were able to identify the three parameters ($b_2$, $b_s$, and SN${}_2$) most responsible for the downward shift. We demonstrated how changes in $P_{\ell}(k)$ due to varying $b_2$, $b_s$, and SN${}_2$ can be ``counteracted'' by changing the shape of the linear power spectrum through the $\Lambda$CDM parameters $\Omega_M$ and $H_0$. Thus, the $\Lambda$CDM fit can explore a larger volume of parameter space for models with low $f\sigma_8$ without significant $\chi^2$ penalties, resulting in a downward shift in marginal constraint from the maximum-likelihood region.  However, we also show that this shift is reduced when the error bars in the data (particularly the low-$k$ part of the quadrupole) shrink via rescalings of the covariance matrix. This suggests that future for surveys, such as DESI, that can provide tighter constraints on $P_\ell(k)$ at large scales this volume effect will not be as significant a concern.

In this paper, we have seen that ``template'' fits of spectroscopic survey data wherein cosmological information is compressed into a small set of physical observables --- in this case the anisotropic clustering amplitude $f\sigma_8$ and the physical size of baryon acoustic oscillations along and perpendicular to the line of sight --- do not (by design!) capture the full phenomenology specific cosmological models (e.g. $\Lambda$CDM) imprint into galaxy clustering. In particular, models like $\Lambda$CDM predict not only the Friedman metric and structure formation at low redshifts but also the high-redshift initial conditions set by pre-decoupling physics, such that parts of parameter space allowed when considering geometry alone can be excluded by this extra information. Within $\Lambda$CDM, for example, we showed that the Alcock-Paczynski distortion $\epsilon$ is strongly constrained to be small not due to the AP effect itself but because the same parameter ($\Omega_M$) that governs it also affects the power spectrum shape. This implies that the template fit is not always an optimal compression of galaxy clustering data even for models with as many cosmological parameters as the compressed variables. For physical models with more degrees of freedom the compressed parameters will be less constrained than in $\Lambda$CDM but, on the other hand, the phenomenology not captured by changes in total amplitude or dilations of the BAO will be richer. We also tested an extension of the template-fit methodology, ``ShapeFit'', showing that like the standard template fit it does not capture the full flexibility of the linear power spectrum shape when compared to direct fits of cosmological parameters, and consequently is not subject to the same volume effects.

The ``template'' fitting method was originally conceived as a way to test the consistency of late-universe large-scale structure data with $\Lambda$CDM at a time when the most stringent constraints on early-universe physics and power spectrum shape came from the CMB. This was sufficient so long as the goal was not to yield constraints on fundamental parameters competitive with, and independent of, the CMB. Looking forward, we expect constraints from galaxy clustering to improve significantly in the coming years. Constraints on the shape of the power spectrum from spectroscopic surveys like BOSS \cite{Dawson13} are already somewhat competitive with the CMB, and future surveys \cite{DESI,Takada14,Schlegel22} will allow us to go further still. As measurements of the linear power spectrum shape become an increasingly important part of the power of spectroscopic surveys, it will be important to incorporate the full physical implication of cosmological models on galaxy clustering beyond late-time effects so that our constraints reflect physically interesting parameter spaces.

\section*{Acknowledgements}
We thank M.~Ivanov, H.~Gil-Marin, and S.~Brieden for comments on an early draft.  S.C.~is supported by the Bezos Membership at the Institute for Advanced Study.  MM and MW are supported by the DOE.
This research has made use of NASA's Astrophysics Data System and the arXiv preprint server.
This research used resources of the National Energy Research Scientific Computing Center (NERSC), a U.S. Department of Energy Office of Science User Facility operated under Contract No.\ DE-AC02-05CH11231.

\bibliography{main}
\bibliographystyle{jhep}

\end{document}